\documentclass[aps,prl,twocolumn, superscriptaddress]{revtex4-1}

\usepackage{amssymb,graphicx,color,amsmath,xfrac,bm,stmaryrd,trimclip} 

\usepackage[mathletters]{ucs}
\usepackage[utf8x]{inputenc}

\definecolor{lightblue}{rgb}{0.17,0.39,1}
\usepackage[bookmarks, colorlinks=true, breaklinks]{hyperref}
\hypersetup{linkcolor=lightblue, citecolor=lightblue, filecolor=black, urlcolor=lightblue}

\begin{document} 

\title{{Homes' Law and Universal Planckian Relaxation}}

\author{A.~Shekhter }
\email[email: ]{arkadyshekhter@gmail.com}
\affiliation{Los Alamos National Laboratory, Los Alamos, NM 87545, USA}

\author{ M.~K.~Chan }
\affiliation{Los Alamos National Laboratory, Los Alamos, NM 87545, USA}

\author{ R.~D.~McDonald }
\affiliation{Los Alamos National Laboratory, Los Alamos, NM 87545, USA}

\author{N.~Harrison}
\affiliation{Los Alamos National Laboratory, Los Alamos, NM 87545, USA}

\begin{abstract} 
According to Zaanen's interpretation of Homes' empirical law~[Zaanen, {\it Nature} {\bf 430}, 512 (2004)], the superconducting transition temperatures in the cuprates are high because their metallic states are as viscous as quantum mechanics permits. Here, we show that Homes' law in fact implies three key points: (i) the resistivity is linear in temperature in the normal state near the transition temperature; (ii) the dimensionless coefficient of proportionality of the relaxation rate with temperature is of order unity---the so-called universal Planckian relaxation rate; and (iii) the logarithmically broad applicability of this law arises from an unusually wide range of effective masses throughout the cuprate phase diagram. In fact, a universal Planckian relaxation rate implies Homes' law only if the mechanism of mass renormalization is independent of the Planckian relaxation.
\end{abstract} 
\date{\today}	
\maketitle



In the high-temperature superconducting cuprates and strongly coupled conventional superconductors, the superfluid density scales with the product of the superconducting transition temperature and the electrical conductivity just above it---a relationship known as ``{Homes' law}''~\cite{homes2004}. Shortly after this empirical observation, Zaanen noted that Homes' law connects two previously unrelated phenomena: the maximum achievable superconducting transition temperatures and a ``strange metal'' state~\cite{anderson1988} that is ``as viscous is permitted by the laws of quantum physics~\cite{zaanen2004}.'' Since then, Homes' law has been shown to apply across a broader class of unconventional superconductors, as well as strongly coupled conventional superconductors~\cite{homes2005,dordevic2013}.

Here, we point out that Homes' law is, in fact, a simple manifestation of the linear-in-temperature resistivity ($\rho = AT$) in the strange metal, which is its defining characteristic. This is understood as a $T$-linear dependence of the scattering rate, $\hbar/\tau = \alpha k_{\rm B} T$, where $\alpha$ is a universal dimensionless quantity of order unity~\cite{legros2019,shekhter2024}. In the cuprates, this relation implies a universal `Planckian' relaxation rate associated with the strange metal state, independent of doping. The law spans many orders of magnitude due to the enormous variation in effective mass~\cite{legros2019,shekhter2024}. Although some of the cuprate sdata used by Homes and coworkers~\cite{homes2004} lie outside the strange metal regime, the law still holds to logarithmic accuracy.

\begin{figure}[!t] 
\begin{center}
\includegraphics[width=0.8\linewidth]{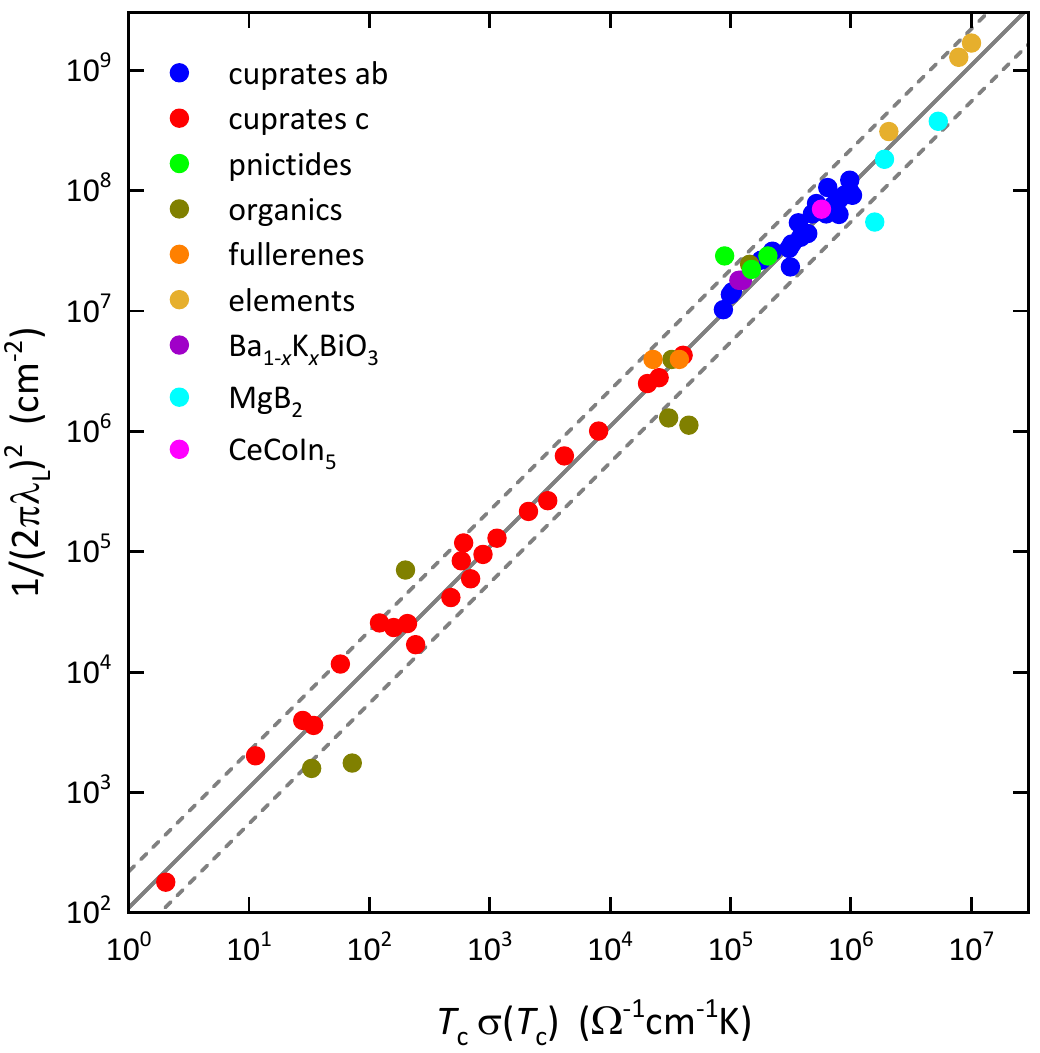}
\textsf{\caption{
{Homes' law plot including data are from Refs.~\onlinecite{homes2004,homes2005,dordevic2013}, with the solid line corresponding to $\alpha=$~2.6 in the expression for $C$. The vertical axis, $(1/2\pi\lambda_{\rm L})^2$, is identical to Refs.~\cite{homes2004,homes2005,dordevic2013}, where this quantity is labeled $\rho_s$, which differs from the standard definition of the superfluid density, $\rho_s = (c/\lambda_{\rm L})^2$, by a factor of $(1/2\pi c)^2$. 
}}
\label{homesplot}}
\end{center}
\vspace{-0.2cm}
\end{figure}

Homes' law states that the zero-temperature superfluid density, $\rho_{\rm s}$, is proportional to the product of the electrical conductivity, $\sigma(T_{\rm c})$, just above the superconducting transition temperature and the transition temperature itself, $T_{\rm c}$. Thus~\cite{homes2004,homes2005,dordevic2013},
\begin{align}\label{homeslaw}
\rho_{\rm s} \propto T_{\rm c}\, \sigma(T_{\rm c}).
\end{align}
The line drawn through the data in Fig.~\ref{homesplot} corresponds specifically to the relation $(1 / 2\pi\lambda_{\rm L})^2 = C\,T_{\rm c}\,\sigma(T_{\rm c})$, where $C$ is the slope. To estimate the slope, we start from the Bardeen-Cooper-Shrieffer (BCS) approximation of the inverse square of the London penetration  depth~\cite{tinkham2004}, given by $\lambda_{\rm L}^{-2} = 4\pi e^2 (n / m^\ast)/c^2$, where $c$ is the speed of light, $e$ is the electron charge, $4\pi e^2(n/m^\ast)$ is the plasma frequency squared, and $n$ is an effective density~\cite{basov2011}.

On the right-hand side of Equation~(\ref{homeslaw}), we adopt the Drude approximation for the conductivity~\cite{ashcroft1976}: $\sigma(T) = e^2 (n / m^\ast) \tau(T)$. Substituting the expression for the Planckian relaxation rate at $T = T_{\rm c}$, $\hbar / \tau = \alpha k_{\rm B} T_{\rm c}$, we arrive at $C = \alpha \times(k_{\rm B} / \hbar) / \pi c^2 \approx \alpha \times 42~\Omega\,\text{cm}^{-1}\,\text{K}^{-1}$ (noting that $8.988×10^{12}$~cm$\,$s$^{-1}$~$= 1\,\Omega^{-1}$, in c.g.s. units). 
Figure~\ref{homesplot} shows a plot of $(1 / 2\pi\lambda_{\rm L})^2$ versus $T_{\rm c}\,\sigma(T_{\rm c})$, with the solid line corresponding to $\alpha = 2.6$, and the dashed lines representing factors of two variation in $\alpha$ above and below this value.

Both the superfluid density and the conductivity are proportional to the ratio $n/m^\ast$—the carrier density divided by the effective mass~\cite{homes2004}. Thus, Homes' law tacitly implies that, in materials where it applies, the relaxation time $\tau$ entering the conductivity is inversely proportional to temperature, $T$. Based solely on dimensional analysis, Zaanen argued~\cite{zaanen2004} that $\hbar/\tau \sim k_{\rm B} T$ at $T = T_{\rm c}$, where $k_{\rm B}$ Boltzmann's constant  and $\hbar$ is Planck's constant (divided by $2\pi$). Since then, $\hbar/\tau \sim k_{\rm B} T$ has come to be known as the Planckian relaxation (or dissipation) rate~\cite{hartnoll2022,phillips2022}, and is believed to apply at all temperatures, not just at $T_{\rm c}$. Empirically, Homes' law yields $\alpha \approx 2.6$, consistent with specific heat, optical conductivity and resistivity measurements~\cite{shekhter2024}.

Homes' law holds with logarithmic accuracy. In the cuprates, for example, the relation $\hbar/\tau = \alpha\, k_{\rm B} T$ accurately describes the resistivity, $\rho(T) = 1/\sigma(T)$, in the strange metal state, which occupies a funnel-shaped region in the temperature–doping phase diagram~\cite{shekhter2024,ando2004}. With the exception of a range of dopings near a critical point $p^\ast \sim 0.2$, the superconducting transition temperature $T_{\rm c}$ lies below the funnel region, meaning that the resistivity in the immediate vicinity of $T_{\rm c}$ departs from the Planckian form. However, because the deviation from $T$-linear resistivity just below the funnel is typically no more than a numerical factor of three, this has little effect on the logarithmic scale in Fig.~\ref{homesplot}.

Given that both the superfluid density and electrical conductivity depend on the ratio $n/m^\ast$, the variation in the plasma frequency squared (rather than the relaxation rate itself) is the primary reason why Homes' law spans multiple decades in Fig.~\ref{homesplot}. In the cuprates, for example, $n/m^\ast$ within the $a,b$-planes varies by roughly an order of magnitude as a function of doping~\cite{legros2019,shekhter2024}. Although both the superfluid density and conductivity are strongly anisotropic in the cuprates~\cite{homes2004}, Homes' law continues to hold because these anisotropies can be attributed to an anisotropy in the effective mass $m^\ast$.It is important to note that $m^\ast$ is much larger in heavy-fermion systems~\cite{hewson1993} than in conventional superconductors~\cite{carbotte1990,bruin2013}. 

Thus, Homes' law clearly demonstrates that superconductivity emerges from an instability in strange metals that occurs irrespective of the underlying effective mass and the mechanism by which it is enhanced. From this perspective, the wide applicability of Homes' law—across a logarithmically broad range of superfluid densities, conductivities, and transition temperatures—suggests that the fluctuations responsible for the linear-in-temperature relaxation rate are largely independent of the electronic correlations that enhance the effective mass. This perspective is discussed in more detail in Ref.~\cite{shekhter2024}.


In the cuprates, the $T$-linear relaxation rate is accompanied by a linear-in-energy relaxation rate, which arises from a competition between frequency and temperature scales associated with quantum criticality~\cite{valla2000,kaminski2005,bok2016,basov2005,armitage2018}. By contrast, phonon fluctuations in conventional metals do not exhibit such competition: while the phonon-scattering relaxation rate depends on temperature, it shows little dependence on energy~\cite{ziman1960}. The fact that the values of $\alpha$ in strongly coupled conventional superconductors are close to unity is attributed to strong charge screening in these systems~\cite{carbotte1990}. In heavy-fermion systems (e.g., CeCoIn$_5$), the Planckian relaxation rate similarly originates from quantum critical fluctuations, as in the cuprates~\cite{vonLoehneysen2007,gegenwart2008,monthoux2007}.

%
%

What sets $\alpha$ close to unity in strange metals? One possibility is that $\alpha \sim 1$ is simply the natural value for such a dimensionless coefficient. Another possibility, as suggested in Ref.~\onlinecite{zaanen2019}, is that $\alpha \sim 1$ represents the maximum allowable value in a system with extensive ``many-body entanglement in the
vacuum.'' Circumstantial evidence for this scenario has been reported in viscosity studies of quark–gluon plasmas~\cite{busza2018} and in the unitary regime of cold-atom Fermi gases~\cite{cao2011}.


Because $\hbar/\tau = \alpha k_{\rm B} T_{\rm c}$ sets the magnitude of the normal-state conductivity at $T_{\rm c}$, Homes' law does not itself impose a bound on $T_{\rm c}$. While a Planckian relaxation rate in the normal state appears to be a prerequisite for high values of the superconducting transition temperature~\cite{zaanen2004}, it does not, in fact, provide an absolute upper limit on $T_{\rm c}$, as originally suggested by Zaanen~\cite{zaanen2004}. In other words, the scale invariance of the Planckian relaxation rate cannot by itself determine the value of $T_{\rm c}$. If anything, it is the magnitude of the plasma frequency that sets the scale of $T_{\rm c}$ in strange metal systems~\cite{emery1995,baskaran1987}. In this sense, the high $T_{\rm c}$ observed in the materials obeying Homes' law is simply a consequence of the fact that strange metals—due to their highly entangled many-body state—become unstable to superconductivity at atypically high temperatures~\cite{zaanen_private_2018,zaanen2004}.

\begin{acknowledgments} {\it Acknowledgements}. This work was performed as part of the Department of Energy (DoE) BES project `Science of 100 tesla.' 
\end{acknowledgments}

\bibliographystyle{apsrev4-1}
\bibliography{homes.bib}

\begin{thebibliography}{31}%
\makeatletter
\providecommand \@ifxundefined [1]{%
 \@ifx{#1\undefined}
}%
\providecommand \@ifnum [1]{%
 \ifnum #1\expandafter \@firstoftwo
 \else \expandafter \@secondoftwo
 \fi
}%
\providecommand \@ifx [1]{%
 \ifx #1\expandafter \@firstoftwo
 \else \expandafter \@secondoftwo
 \fi
}%
\providecommand \natexlab [1]{#1}%
\providecommand \enquote  [1]{``#1''}%
\providecommand \bibnamefont  [1]{#1}%
\providecommand \bibfnamefont [1]{#1}%
\providecommand \citenamefont [1]{#1}%
\providecommand \href@noop [0]{\@secondoftwo}%
\providecommand \href [0]{\begingroup \@sanitize@url \@href}%
\providecommand \@href[1]{\@@startlink{#1}\@@href}%
\providecommand \@@href[1]{\endgroup#1\@@endlink}%
\providecommand \@sanitize@url [0]{\catcode `\\12\catcode `\$12\catcode
  `\&12\catcode `\#12\catcode `\^12\catcode `\_12\catcode `\%12\relax}%
\providecommand \@@startlink[1]{}%
\providecommand \@@endlink[0]{}%
\providecommand \url  [0]{\begingroup\@sanitize@url \@url }%
\providecommand \@url [1]{\endgroup\@href {#1}{\urlprefix }}%
\providecommand \urlprefix  [0]{URL }%
\providecommand \Eprint [0]{\href }%
\providecommand \doibase [0]{http://dx.doi.org/}%
\providecommand \selectlanguage [0]{\@gobble}%
\providecommand \bibinfo  [0]{\@secondoftwo}%
\providecommand \bibfield  [0]{\@secondoftwo}%
\providecommand \translation [1]{[#1]}%
\providecommand \BibitemOpen [0]{}%
\providecommand \bibitemStop [0]{}%
\providecommand \bibitemNoStop [0]{.\EOS\space}%
\providecommand \EOS [0]{\spacefactor3000\relax}%
\providecommand \BibitemShut  [1]{\csname bibitem#1\endcsname}%
\let\auto@bib@innerbib\@empty
\bibitem [{\citenamefont {Homes}\ \emph {et~al.}(2004)\citenamefont {Homes},
  \citenamefont {Dordevic}, \citenamefont {Strongin}, \citenamefont {Bonn},
  \citenamefont {Liang}, \citenamefont {Hardy}, \citenamefont {Komiya},
  \citenamefont {Ando}, \citenamefont {Yu}, \citenamefont {Kaneko},
  \citenamefont {Zhao}, \citenamefont {Greven}, \citenamefont {Basov},\ and\
  \citenamefont {Timusk}}]{homes2004}%
  \BibitemOpen
  \bibfield  {author} {\bibinfo {author} {\bibfnamefont {C.~C.}\ \bibnamefont
  {Homes}}, \bibinfo {author} {\bibfnamefont {S.~V.}\ \bibnamefont {Dordevic}},
  \bibinfo {author} {\bibfnamefont {M.}~\bibnamefont {Strongin}}, \bibinfo
  {author} {\bibfnamefont {D.~A.}\ \bibnamefont {Bonn}}, \bibinfo {author}
  {\bibfnamefont {R.}~\bibnamefont {Liang}}, \bibinfo {author} {\bibfnamefont
  {W.~N.}\ \bibnamefont {Hardy}}, \bibinfo {author} {\bibfnamefont
  {S.}~\bibnamefont {Komiya}}, \bibinfo {author} {\bibfnamefont
  {Y.}~\bibnamefont {Ando}}, \bibinfo {author} {\bibfnamefont {G.}~\bibnamefont
  {Yu}}, \bibinfo {author} {\bibfnamefont {N.}~\bibnamefont {Kaneko}}, \bibinfo
  {author} {\bibfnamefont {X.}~\bibnamefont {Zhao}}, \bibinfo {author}
  {\bibfnamefont {M.}~\bibnamefont {Greven}}, \bibinfo {author} {\bibfnamefont
  {D.~N.}\ \bibnamefont {Basov}}, \ and\ \bibinfo {author} {\bibfnamefont
  {T.}~\bibnamefont {Timusk}},\ }\href {\doibase 10.1038/nature02773}
  {\bibfield  {journal} {\bibinfo  {journal} {Nature}\ }\textbf {\bibinfo
  {volume} {430}},\ \bibinfo {pages} {539} (\bibinfo {year}
  {2004})}\BibitemShut {NoStop}%
\bibitem [{\citenamefont {Anderson}(1988)}]{anderson1988}%
  \BibitemOpen
  \bibfield  {author} {\bibinfo {author} {\bibfnamefont {P.~W.}\ \bibnamefont
  {Anderson}},\ }in\ \href {\doibase 10.1063/1.37198} {\emph {\bibinfo
  {booktitle} {AIP Conference Proceedings}}},\ Vol.\ \bibinfo {volume} {169}\
  (\bibinfo  {publisher} {AIP Publishing},\ \bibinfo {year} {1988})\ pp.\
  \bibinfo {pages} {141--157}\BibitemShut {NoStop}%
\bibitem [{\citenamefont {Zaanen}(2004)}]{zaanen2004}%
  \BibitemOpen
  \bibfield  {author} {\bibinfo {author} {\bibfnamefont {J.}~\bibnamefont
  {Zaanen}},\ }\href {\doibase 10.1038/430512a} {\bibfield  {journal} {\bibinfo
   {journal} {Nature}\ }\textbf {\bibinfo {volume} {430}},\ \bibinfo {pages}
  {512} (\bibinfo {year} {2004})}\BibitemShut {NoStop}%
\bibitem [{\citenamefont {Homes}\ \emph {et~al.}(2005)\citenamefont {Homes},
  \citenamefont {Dordevic}, \citenamefont {Valla},\ and\ \citenamefont
  {Strongin}}]{homes2005}%
  \BibitemOpen
  \bibfield  {author} {\bibinfo {author} {\bibfnamefont {C.~C.}\ \bibnamefont
  {Homes}}, \bibinfo {author} {\bibfnamefont {S.~V.}\ \bibnamefont {Dordevic}},
  \bibinfo {author} {\bibfnamefont {T.}~\bibnamefont {Valla}}, \ and\ \bibinfo
  {author} {\bibfnamefont {M.}~\bibnamefont {Strongin}},\ }\href {\doibase
  10.1103/PhysRevB.72.134517} {\bibfield  {journal} {\bibinfo  {journal}
  {Physical Review B}\ }\textbf {\bibinfo {volume} {72}},\ \bibinfo {pages}
  {134517} (\bibinfo {year} {2005})}\BibitemShut {NoStop}%
\bibitem [{\citenamefont {Dordevic}\ \emph {et~al.}(2013)\citenamefont
  {Dordevic}, \citenamefont {Basov},\ and\ \citenamefont
  {Homes}}]{dordevic2013}%
  \BibitemOpen
  \bibfield  {author} {\bibinfo {author} {\bibfnamefont {S.~V.}\ \bibnamefont
  {Dordevic}}, \bibinfo {author} {\bibfnamefont {D.~N.}\ \bibnamefont {Basov}},
  \ and\ \bibinfo {author} {\bibfnamefont {C.~C.}\ \bibnamefont {Homes}},\
  }\href {\doibase 10.1038/srep01713} {\bibfield  {journal} {\bibinfo
  {journal} {Scientific Reports}\ }\textbf {\bibinfo {volume} {3}},\ \bibinfo
  {pages} {1713} (\bibinfo {year} {2013})}\BibitemShut {NoStop}%
\bibitem [{\citenamefont {Legros}\ \emph {et~al.}(2019)\citenamefont {Legros}
  \emph {et~al.}}]{legros2019}%
  \BibitemOpen
  \bibfield  {author} {\bibinfo {author} {\bibfnamefont {A.}~\bibnamefont
  {Legros}} \emph {et~al.},\ }\href {\doibase 10.1038/s41567-018-0334-2}
  {\bibfield  {journal} {\bibinfo  {journal} {Nature Phys.}\ }\textbf {\bibinfo
  {volume} {15}},\ \bibinfo {pages} {142} (\bibinfo {year} {2019})}\BibitemShut
  {NoStop}%
\bibitem [{\citenamefont {Shekhter}\ \emph {et~al.}(2024)\citenamefont
  {Shekhter}, \citenamefont {Ramshaw}, \citenamefont {Chan},\ and\
  \citenamefont {Harrison}}]{shekhter2024}%
  \BibitemOpen
  \bibfield  {author} {\bibinfo {author} {\bibfnamefont {A.}~\bibnamefont
  {Shekhter}}, \bibinfo {author} {\bibfnamefont {B.~J.}\ \bibnamefont
  {Ramshaw}}, \bibinfo {author} {\bibfnamefont {M.~K.}\ \bibnamefont {Chan}}, \
  and\ \bibinfo {author} {\bibfnamefont {N.}~\bibnamefont {Harrison}},\ }\href
  {https://arxiv.org/abs/2406.12133} {\bibfield  {journal} {\bibinfo  {journal}
  {arXiv preprint arXiv:2406.12133}\ } (\bibinfo {year} {2024})},\ \Eprint
  {http://arxiv.org/abs/2406.12133} {arXiv:2406.12133 [cond-mat.str-el]}
  \BibitemShut {NoStop}%
\bibitem [{\citenamefont {Tinkham}(2004)}]{tinkham2004}%
  \BibitemOpen
  \bibfield  {author} {\bibinfo {author} {\bibfnamefont {M.}~\bibnamefont
  {Tinkham}},\ }\href@noop {} {\emph {\bibinfo {title} {Introduction to
  Superconductivity}}},\ \bibinfo {edition} {2nd}\ ed.\ (\bibinfo  {publisher}
  {Dover Publications},\ \bibinfo {address} {Mineola, New York},\ \bibinfo
  {year} {2004})\BibitemShut {NoStop}%
\bibitem [{\citenamefont {Basov}\ \emph {et~al.}(2011)\citenamefont {Basov},
  \citenamefont {Averitt}, \citenamefont {van~der Marel}, \citenamefont
  {Dressel},\ and\ \citenamefont {Haule}}]{basov2011}%
  \BibitemOpen
  \bibfield  {author} {\bibinfo {author} {\bibfnamefont {D.~N.}\ \bibnamefont
  {Basov}}, \bibinfo {author} {\bibfnamefont {R.~D.}\ \bibnamefont {Averitt}},
  \bibinfo {author} {\bibfnamefont {D.}~\bibnamefont {van~der Marel}}, \bibinfo
  {author} {\bibfnamefont {M.}~\bibnamefont {Dressel}}, \ and\ \bibinfo
  {author} {\bibfnamefont {K.}~\bibnamefont {Haule}},\ }\href {\doibase
  10.1103/RevModPhys.83.471} {\bibfield  {journal} {\bibinfo  {journal} {Rev.
  Mod. Phys.}\ }\textbf {\bibinfo {volume} {83}},\ \bibinfo {pages} {471}
  (\bibinfo {year} {2011})}\BibitemShut {NoStop}%
\bibitem [{\citenamefont {Ashcroft}\ and\ \citenamefont
  {Mermin}(1976)}]{ashcroft1976}%
  \BibitemOpen
  \bibfield  {author} {\bibinfo {author} {\bibfnamefont {N.~W.}\ \bibnamefont
  {Ashcroft}}\ and\ \bibinfo {author} {\bibfnamefont {N.~D.}\ \bibnamefont
  {Mermin}},\ }\href@noop {} {\emph {\bibinfo {title} {Solid State Physics}}}\
  (\bibinfo  {publisher} {Saunders College Publishing},\ \bibinfo {address}
  {Orlando},\ \bibinfo {year} {1976})\BibitemShut {NoStop}%
\bibitem [{\citenamefont {Hartnoll}\ and\ \citenamefont
  {Mackenzie}(2022)}]{hartnoll2022}%
  \BibitemOpen
  \bibfield  {author} {\bibinfo {author} {\bibfnamefont {S.~A.}\ \bibnamefont
  {Hartnoll}}\ and\ \bibinfo {author} {\bibfnamefont {A.~P.}\ \bibnamefont
  {Mackenzie}},\ }\href {\doibase 10.1103/RevModPhys.94.041002} {\bibfield
  {journal} {\bibinfo  {journal} {Reviews of Modern Physics}\ }\textbf
  {\bibinfo {volume} {94}} (\bibinfo {year} {2022}),\
  10.1103/RevModPhys.94.041002}\BibitemShut {NoStop}%
\bibitem [{\citenamefont {Phillips}\ \emph {et~al.}(2022)\citenamefont
  {Phillips}, \citenamefont {Hussey},\ and\ \citenamefont
  {Abbamonte}}]{phillips2022}%
  \BibitemOpen
  \bibfield  {author} {\bibinfo {author} {\bibfnamefont {P.~W.}\ \bibnamefont
  {Phillips}}, \bibinfo {author} {\bibfnamefont {N.~E.}\ \bibnamefont
  {Hussey}}, \ and\ \bibinfo {author} {\bibfnamefont {P.}~\bibnamefont
  {Abbamonte}},\ }\href {\doibase 10.1126/science.abh4273} {\bibfield
  {journal} {\bibinfo  {journal} {Science}\ }\textbf {\bibinfo {volume} {377}}
  (\bibinfo {year} {2022}),\ 10.1126/science.abh4273}\BibitemShut {NoStop}%
\bibitem [{\citenamefont {Ando}\ \emph {et~al.}(2004)\citenamefont {Ando},
  \citenamefont {Komiya}, \citenamefont {Segawa}, \citenamefont {Ono},\ and\
  \citenamefont {Kurita}}]{ando2004}%
  \BibitemOpen
  \bibfield  {author} {\bibinfo {author} {\bibfnamefont {Y.}~\bibnamefont
  {Ando}}, \bibinfo {author} {\bibfnamefont {S.}~\bibnamefont {Komiya}},
  \bibinfo {author} {\bibfnamefont {K.}~\bibnamefont {Segawa}}, \bibinfo
  {author} {\bibfnamefont {S.}~\bibnamefont {Ono}}, \ and\ \bibinfo {author}
  {\bibfnamefont {Y.}~\bibnamefont {Kurita}},\ }\href {\doibase
  10.1103/PhysRevLett.93.267001} {\bibfield  {journal} {\bibinfo  {journal}
  {Phys. Rev. Lett.}\ }\textbf {\bibinfo {volume} {93}},\ \bibinfo {pages}
  {267001} (\bibinfo {year} {2004})}\BibitemShut {NoStop}%
\bibitem [{\citenamefont {Hewson}(1993)}]{hewson1993}%
  \BibitemOpen
  \bibfield  {author} {\bibinfo {author} {\bibfnamefont {A.~C.}\ \bibnamefont
  {Hewson}},\ }\href@noop {} {\emph {\bibinfo {title} {The Kondo Problem to
  Heavy Fermions}}},\ Cambridge Studies in Magnetism\ (\bibinfo  {publisher}
  {Cambridge University Press},\ \bibinfo {address} {Cambridge},\ \bibinfo
  {year} {1993})\BibitemShut {NoStop}%
\bibitem [{\citenamefont {Carbotte}(1990)}]{carbotte1990}%
  \BibitemOpen
  \bibfield  {author} {\bibinfo {author} {\bibfnamefont {J.~P.}\ \bibnamefont
  {Carbotte}},\ }\href {\doibase 10.1103/RevModPhys.62.1027} {\bibfield
  {journal} {\bibinfo  {journal} {Reviews of Modern Physics}\ }\textbf
  {\bibinfo {volume} {62}},\ \bibinfo {pages} {1027} (\bibinfo {year}
  {1990})}\BibitemShut {NoStop}%
\bibitem [{\citenamefont {Bruin}\ \emph {et~al.}(2013)\citenamefont {Bruin},
  \citenamefont {Sakai}, \citenamefont {Perry},\ and\ \citenamefont
  {Mackenzie}}]{bruin2013}%
  \BibitemOpen
  \bibfield  {author} {\bibinfo {author} {\bibfnamefont {J.~A.~N.}\
  \bibnamefont {Bruin}}, \bibinfo {author} {\bibfnamefont {H.}~\bibnamefont
  {Sakai}}, \bibinfo {author} {\bibfnamefont {R.~S.}\ \bibnamefont {Perry}}, \
  and\ \bibinfo {author} {\bibfnamefont {A.~P.}\ \bibnamefont {Mackenzie}},\
  }\href {\doibase 10.1126/science.1227612} {\bibfield  {journal} {\bibinfo
  {journal} {Science}\ }\textbf {\bibinfo {volume} {339}},\ \bibinfo {pages}
  {804} (\bibinfo {year} {2013})}\BibitemShut {NoStop}%
\bibitem [{\citenamefont {Valla}\ \emph {et~al.}(2000)\citenamefont {Valla},
  \citenamefont {Fedorov}, \citenamefont {Johnson}, \citenamefont {Li},
  \citenamefont {Gu},\ and\ \citenamefont {Koshizuka}}]{valla2000}%
  \BibitemOpen
  \bibfield  {author} {\bibinfo {author} {\bibfnamefont {T.}~\bibnamefont
  {Valla}}, \bibinfo {author} {\bibfnamefont {A.~V.}\ \bibnamefont {Fedorov}},
  \bibinfo {author} {\bibfnamefont {P.~D.}\ \bibnamefont {Johnson}}, \bibinfo
  {author} {\bibfnamefont {Q.}~\bibnamefont {Li}}, \bibinfo {author}
  {\bibfnamefont {G.~D.}\ \bibnamefont {Gu}}, \ and\ \bibinfo {author}
  {\bibfnamefont {N.}~\bibnamefont {Koshizuka}},\ }\href {\doibase
  10.1103/PhysRevLett.85.828} {\bibfield  {journal} {\bibinfo  {journal} {Phys.
  Rev. Lett.}\ }\textbf {\bibinfo {volume} {85}},\ \bibinfo {pages} {828}
  (\bibinfo {year} {2000})}\BibitemShut {NoStop}%
\bibitem [{\citenamefont {Kaminski}\ \emph {et~al.}(2005)\citenamefont
  {Kaminski}, \citenamefont {Fretwell}, \citenamefont {Norman}, \citenamefont
  {Randeria}, \citenamefont {Rosenkranz}, \citenamefont {Chatterjee},
  \citenamefont {Campuzano}, \citenamefont {Mesot}, \citenamefont {Sato},
  \citenamefont {Takahashi}, \citenamefont {Terashima}, \citenamefont {Takano},
  \citenamefont {Kadowaki}, \citenamefont {Li},\ and\ \citenamefont
  {Raffy}}]{kaminski2005}%
  \BibitemOpen
  \bibfield  {author} {\bibinfo {author} {\bibfnamefont {A.}~\bibnamefont
  {Kaminski}}, \bibinfo {author} {\bibfnamefont {H.~M.}\ \bibnamefont
  {Fretwell}}, \bibinfo {author} {\bibfnamefont {M.~R.}\ \bibnamefont
  {Norman}}, \bibinfo {author} {\bibfnamefont {M.}~\bibnamefont {Randeria}},
  \bibinfo {author} {\bibfnamefont {S.}~\bibnamefont {Rosenkranz}}, \bibinfo
  {author} {\bibfnamefont {U.}~\bibnamefont {Chatterjee}}, \bibinfo {author}
  {\bibfnamefont {J.~C.}\ \bibnamefont {Campuzano}}, \bibinfo {author}
  {\bibfnamefont {J.}~\bibnamefont {Mesot}}, \bibinfo {author} {\bibfnamefont
  {T.}~\bibnamefont {Sato}}, \bibinfo {author} {\bibfnamefont {T.}~\bibnamefont
  {Takahashi}}, \bibinfo {author} {\bibfnamefont {T.}~\bibnamefont
  {Terashima}}, \bibinfo {author} {\bibfnamefont {M.}~\bibnamefont {Takano}},
  \bibinfo {author} {\bibfnamefont {K.}~\bibnamefont {Kadowaki}}, \bibinfo
  {author} {\bibfnamefont {Z.~Z.}\ \bibnamefont {Li}}, \ and\ \bibinfo {author}
  {\bibfnamefont {H.}~\bibnamefont {Raffy}},\ }\href {\doibase
  10.1103/PhysRevB.71.014517} {\bibfield  {journal} {\bibinfo  {journal} {Phys.
  Rev. B}\ }\textbf {\bibinfo {volume} {71}},\ \bibinfo {pages} {014517}
  (\bibinfo {year} {2005})}\BibitemShut {NoStop}%
\bibitem [{\citenamefont {Bok}\ \emph {et~al.}(2016)\citenamefont {Bok},
  \citenamefont {Bae}, \citenamefont {Choi}, \citenamefont {Varma},
  \citenamefont {Zhang}, \citenamefont {He}, \citenamefont {Zhang},
  \citenamefont {Yu},\ and\ \citenamefont {Zhou}}]{bok2016}%
  \BibitemOpen
  \bibfield  {author} {\bibinfo {author} {\bibfnamefont {J.~M.}\ \bibnamefont
  {Bok}}, \bibinfo {author} {\bibfnamefont {J.~J.}\ \bibnamefont {Bae}},
  \bibinfo {author} {\bibfnamefont {H.-Y.}\ \bibnamefont {Choi}}, \bibinfo
  {author} {\bibfnamefont {C.~M.}\ \bibnamefont {Varma}}, \bibinfo {author}
  {\bibfnamefont {W.}~\bibnamefont {Zhang}}, \bibinfo {author} {\bibfnamefont
  {J.}~\bibnamefont {He}}, \bibinfo {author} {\bibfnamefont {Y.}~\bibnamefont
  {Zhang}}, \bibinfo {author} {\bibfnamefont {L.}~\bibnamefont {Yu}}, \ and\
  \bibinfo {author} {\bibfnamefont {X.~J.}\ \bibnamefont {Zhou}},\ }\href
  {\doibase 10.1126/sciadv.1501329} {\bibfield  {journal} {\bibinfo  {journal}
  {Science Advances}\ }\textbf {\bibinfo {volume} {2}},\ \bibinfo {pages}
  {e1501329} (\bibinfo {year} {2016})}\BibitemShut {NoStop}%
\bibitem [{\citenamefont {Basov}\ and\ \citenamefont
  {Timusk}(2005)}]{basov2005}%
  \BibitemOpen
  \bibfield  {author} {\bibinfo {author} {\bibfnamefont {D.~N.}\ \bibnamefont
  {Basov}}\ and\ \bibinfo {author} {\bibfnamefont {T.}~\bibnamefont {Timusk}},\
  }\href {\doibase 10.1103/RevModPhys.77.721} {\bibfield  {journal} {\bibinfo
  {journal} {Rev. Mod. Phys.}\ }\textbf {\bibinfo {volume} {77}},\ \bibinfo
  {pages} {721} (\bibinfo {year} {2005})}\BibitemShut {NoStop}%
\bibitem [{\citenamefont {Armitage}(2018)}]{armitage2018}%
  \BibitemOpen
  \bibfield  {author} {\bibinfo {author} {\bibfnamefont {N.}~\bibnamefont
  {Armitage}},\ }\href@noop {} {\enquote {\bibinfo {title} {Electrodynamics of
  correlated electron systems},}\ } (\bibinfo {year} {2018}),\ \Eprint
  {http://arxiv.org/abs/0908.1126} {arXiv:0908.1126 [cond-mat.str-el]}
  \BibitemShut {NoStop}%
\bibitem [{\citenamefont {Ziman}(1960)}]{ziman1960}%
  \BibitemOpen
  \bibfield  {author} {\bibinfo {author} {\bibfnamefont {J.}~\bibnamefont
  {Ziman}},\ }\href@noop {} {\emph {\bibinfo {title} {Electrons and Phonons:
  The Theory of Transport Phenomena in Solids}}},\ Oxford Classic Texts in the
  Physical Sciences\ (\bibinfo  {publisher} {Oxford University Press},\
  \bibinfo {address} {Oxford},\ \bibinfo {year} {1960})\BibitemShut {NoStop}%
\bibitem [{\citenamefont {von L\"ohneysen}\ \emph {et~al.}(2007)\citenamefont
  {von L\"ohneysen}, \citenamefont {Rosch}, \citenamefont {Vojta},\ and\
  \citenamefont {W\"olfle}}]{vonLoehneysen2007}%
  \BibitemOpen
  \bibfield  {author} {\bibinfo {author} {\bibfnamefont {H.}~\bibnamefont {von
  L\"ohneysen}}, \bibinfo {author} {\bibfnamefont {A.}~\bibnamefont {Rosch}},
  \bibinfo {author} {\bibfnamefont {M.}~\bibnamefont {Vojta}}, \ and\ \bibinfo
  {author} {\bibfnamefont {P.}~\bibnamefont {W\"olfle}},\ }\href {\doibase
  10.1103/RevModPhys.79.1015} {\bibfield  {journal} {\bibinfo  {journal}
  {Reviews of Modern Physics}\ }\textbf {\bibinfo {volume} {79}},\ \bibinfo
  {pages} {1015} (\bibinfo {year} {2007})}\BibitemShut {NoStop}%
\bibitem [{\citenamefont {Gegenwart}\ \emph {et~al.}(2008)\citenamefont
  {Gegenwart}, \citenamefont {Si},\ and\ \citenamefont
  {Steglich}}]{gegenwart2008}%
  \BibitemOpen
  \bibfield  {author} {\bibinfo {author} {\bibfnamefont {P.}~\bibnamefont
  {Gegenwart}}, \bibinfo {author} {\bibfnamefont {Q.}~\bibnamefont {Si}}, \
  and\ \bibinfo {author} {\bibfnamefont {F.}~\bibnamefont {Steglich}},\ }\href
  {\doibase 10.1038/nphys892} {\bibfield  {journal} {\bibinfo  {journal} {Nat.
  Phys.}\ }\textbf {\bibinfo {volume} {4}},\ \bibinfo {pages} {186} (\bibinfo
  {year} {2008})}\BibitemShut {NoStop}%
\bibitem [{\citenamefont {Monthoux}\ \emph {et~al.}(2007)\citenamefont
  {Monthoux}, \citenamefont {Pines},\ and\ \citenamefont
  {Lonzarich}}]{monthoux2007}%
  \BibitemOpen
  \bibfield  {author} {\bibinfo {author} {\bibfnamefont {P.}~\bibnamefont
  {Monthoux}}, \bibinfo {author} {\bibfnamefont {D.}~\bibnamefont {Pines}}, \
  and\ \bibinfo {author} {\bibfnamefont {G.~G.}\ \bibnamefont {Lonzarich}},\
  }\href {\doibase 10.1038/nature06480} {\bibfield  {journal} {\bibinfo
  {journal} {Nature}\ }\textbf {\bibinfo {volume} {450}},\ \bibinfo {pages}
  {1177} (\bibinfo {year} {2007})}\BibitemShut {NoStop}%
\bibitem [{\citenamefont {Zaanen}(2019)}]{zaanen2019}%
  \BibitemOpen
  \bibfield  {author} {\bibinfo {author} {\bibfnamefont {J.}~\bibnamefont
  {Zaanen}},\ }\href {\doibase 10.21468/SciPostPhys.6.5.061} {\bibfield
  {journal} {\bibinfo  {journal} {SciPost Phys.}\ }\textbf {\bibinfo {volume}
  {6}},\ \bibinfo {pages} {061} (\bibinfo {year} {2019})}\BibitemShut {NoStop}%
\bibitem [{\citenamefont {Busza}\ \emph {et~al.}(2018)\citenamefont {Busza},
  \citenamefont {Rajagopal},\ and\ \citenamefont {van~der Schee}}]{busza2018}%
  \BibitemOpen
  \bibfield  {author} {\bibinfo {author} {\bibfnamefont {W.}~\bibnamefont
  {Busza}}, \bibinfo {author} {\bibfnamefont {K.}~\bibnamefont {Rajagopal}}, \
  and\ \bibinfo {author} {\bibfnamefont {W.}~\bibnamefont {van~der Schee}},\
  }\href {\doibase 10.1146/annurev-nucl-101917-020852} {\bibfield  {journal}
  {\bibinfo  {journal} {Annual Review of Nuclear and Particle Science}\
  }\textbf {\bibinfo {volume} {68}},\ \bibinfo {pages} {339} (\bibinfo {year}
  {2018})}\BibitemShut {NoStop}%
\bibitem [{\citenamefont {Cao}\ \emph {et~al.}(2011)\citenamefont {Cao},
  \citenamefont {Elliott}, \citenamefont {Joseph}, \citenamefont {Wu},
  \citenamefont {Petricka}, \citenamefont {Sch{\"a}fer},\ and\ \citenamefont
  {Thomas}}]{cao2011}%
  \BibitemOpen
  \bibfield  {author} {\bibinfo {author} {\bibfnamefont {C.}~\bibnamefont
  {Cao}}, \bibinfo {author} {\bibfnamefont {E.}~\bibnamefont {Elliott}},
  \bibinfo {author} {\bibfnamefont {J.}~\bibnamefont {Joseph}}, \bibinfo
  {author} {\bibfnamefont {H.}~\bibnamefont {Wu}}, \bibinfo {author}
  {\bibfnamefont {J.}~\bibnamefont {Petricka}}, \bibinfo {author}
  {\bibfnamefont {T.}~\bibnamefont {Sch{\"a}fer}}, \ and\ \bibinfo {author}
  {\bibfnamefont {J.~E.}\ \bibnamefont {Thomas}},\ }\href {\doibase
  10.1126/science.1195219} {\bibfield  {journal} {\bibinfo  {journal}
  {Science}\ }\textbf {\bibinfo {volume} {331}},\ \bibinfo {pages} {58}
  (\bibinfo {year} {2011})}\BibitemShut {NoStop}%
\bibitem [{\citenamefont {Emery}\ and\ \citenamefont
  {Kivelson}(1995)}]{emery1995}%
  \BibitemOpen
  \bibfield  {author} {\bibinfo {author} {\bibfnamefont {V.}~\bibnamefont
  {Emery}}\ and\ \bibinfo {author} {\bibfnamefont {S.}~\bibnamefont
  {Kivelson}},\ }\href {\doibase 10.1038/374434a0} {\bibfield  {journal}
  {\bibinfo  {journal} {Nature}\ }\textbf {\bibinfo {volume} {374}},\ \bibinfo
  {pages} {434} (\bibinfo {year} {1995})}\BibitemShut {NoStop}%
\bibitem [{\citenamefont {Baskaran}\ \emph {et~al.}(1987)\citenamefont
  {Baskaran}, \citenamefont {Zou},\ and\ \citenamefont
  {Anderson}}]{baskaran1987}%
  \BibitemOpen
  \bibfield  {author} {\bibinfo {author} {\bibfnamefont {G.}~\bibnamefont
  {Baskaran}}, \bibinfo {author} {\bibfnamefont {Z.}~\bibnamefont {Zou}}, \
  and\ \bibinfo {author} {\bibfnamefont {P.~W.}\ \bibnamefont {Anderson}},\
  }\href {\doibase 10.1016/0038-1098(87)90642-9} {\bibfield  {journal}
  {\bibinfo  {journal} {Solid State Communications}\ }\textbf {\bibinfo
  {volume} {63}},\ \bibinfo {pages} {973} (\bibinfo {year} {1987})}\BibitemShut
  {NoStop}%
\bibitem [{\citenamefont {Zaanen}(2018)}]{zaanen_private_2018}%
  \BibitemOpen
  \bibfield  {author} {\bibinfo {author} {\bibfnamefont {J.}~\bibnamefont
  {Zaanen}},\ }\href@noop {} {\enquote {\bibinfo {title} {Private
  communication},}\ } (\bibinfo {year} {2018})\BibitemShut {NoStop}%
\end{thebibliography}%

\end{document}